\documentclass[fleqn,10pt]{wlscirep}
\usepackage[utf8]{inputenc}
\usepackage[T1]{fontenc}
\usepackage{marvosym}
\usepackage{authblk}
\usepackage{xcolor}
\usepackage{soul}

\usepackage{diagbox}
\usepackage{amsmath}
\usepackage{subcaption}
\usepackage{amssymb} 

\title{Quantitative multi-metabolite imaging of Parkinson's disease using AI boosted molecular MRI}

\author[1]{Hagar Shmuely}
\author[1]{Michal Rivlin}
\author[1,2*]{Or Perlman}
\affil[1]{School of Biomedical Engineering, Tel Aviv University, Tel Aviv, Israel}
\affil[2]{Sagol School of Neuroscience, Tel Aviv University, Tel Aviv, Israel}

\affil[*]{Correspondence: orperlman@tauex.tau.ac.il}

\begin{abstract}
Traditional approaches for molecular imaging of Parkinson's disease (PD) in vivo require radioactive isotopes, lengthy scan times, or deliver only low spatial resolution. Recent advances in saturation transfer-based PD magnetic resonance imaging (MRI) have provided biochemical insights, although the image contrast is semi-quantitative and nonspecific. Here, we combined a rapid molecular MRI acquisition paradigm with deep learning based reconstruction for multi-metabolite quantification of glutamate, mobile proteins, semisolid, and mobile macromolecules in an acute MPTP (1-methyl-4-phenyl-1,2,3,6-tetrahydropyridine) mouse model. The quantitative parameter maps are in general agreement with the histology and MR spectroscopy, and demonstrate that semisolid magnetization transfer (MT), amide, and aliphatic relayed nuclear Overhauser effect (rNOE) proton volume fractions may serve as PD biomarkers.

\end{abstract}

\begin{document}

\flushbottom
\maketitle

\thispagestyle{empty}

\section*{Introduction}
Parkinson’s disease (PD), the second-most-common neurodegenerative disorder, is driven by multiple pathological processes\cite{Cramb2023}. Disease hallmarks include misfolded $ \alpha $-sinuclein protein aggregation and dopaminergic neuronal loss (mainly in the substantia nigra, but also in other brain regions, such as the striatum), resulting in dopamine deficiency and dysregulation of the basal ganglia circuitry\cite{Balestrino2019, Poewe2017}. PD progression also involves glial cell dysfunction, with previous studies reporting dramatic striatal reactive astrogliosis (including astrocyte proliferation and volume increase)\cite{Charron2014,Luchtman2009, KurkowskaJastrzbska2009}. This leads to impaired glutamate reuptake and its accumulation in the synapse\cite{Iovino2020}.

As PD diagnosis is commonly based on motor symptom manifestation, which occurs relatively late in the disease timeline\cite{Kordower2013}, there is an urgent need for an alternative, non-invasive approach to detect the molecular phenomena associated with prodromal disease. Accurate characterization of molecular effects in vivo could also improve the differentiation of PD subtypes and stages, thereby introducing new options for a personally customized effective therapeutic regimen\cite{Fereshtehnejad2017}.

% here I kept the original order
Neuroimaging has long been used in PD research\cite{Politis2014} and has demonstrated clinical utility\cite{Morbelli2020}. Positron emission tomography (PET) and single photon emission computed tomography (SPECT) can assess the integrity of dopaminergic neurons and synapses, yet they require radioactive materials and deliver poor spatial resolution\cite{Bidesi2021}. Magnetic resonance imaging (MRI) provides excellent soft tissue contrast without the need for ionizing radiation. However, standard MRI protocols are typically limited for dismissing other possible diagnoses\cite{Tolosa2021, Madelung2025}. Magnetic resonance spectroscopy (MRS), on the other hand, offers metabolic insights, including the quantification of glutamate concentration, yet its low sensitivity translates into long scan times and limited spatial resolution.

Chemical exchange saturation transfer (CEST) MRI is an increasingly investigated contrast mechanism\cite{vinogradov2023cest, rivlin2023metabolic, bricco2023genetic, perlman2020redesigned, mohanta2024intramolecular} that offers improved resolution and faster scan times than MRS\cite{van2011chemical, zaiss2013chemical}. The technique utilizes frequency-selective pulses to saturate the signals of exchangeable protons, which then undergo chemical exchange with water protons, resulting in a decrease in the water signal\cite{vinogradov2013cest}. Initially applied for intracellular pH imaging in stroke\cite{Zhou2003} and following the dynamics of endogenous cellular proteins and peptides in cancer\cite{zhou2011differentiation}, CEST is now recognized as able to detect glutamate-associated signals\cite{cai2012magnetic}. In the context of PD, several CEST studies in murine PD models or human patients have reported an elevated glutamate CEST (GluCEST) signal\cite{bagga2016mapping, Bagga2018, Liu2025}, increased amide proton transfer (APT) CEST contrast\cite{Li2014}, and altered aliphatic relayed nuclear Overhauser (rNOE) signals\cite{Mennecke2022}.

However, while the standard CEST-weighted signal is proportional to the concentration of the compound of interest, it is also affected by the proton exchange rate, water relaxivity, and the particular properties used in the acquisition pulse sequence\cite{van2011chemical}. This complicates the interpretation of the resulting contrast and the derivation of definitive conclusions about the molecular tissue content. Moreover, various CEST targets are characterized by different proton exchange rates, necessitating the use of a distinct set of acquisition parameters to be applied as part of a separate acquisition protocol. For example, a much higher saturation power is required to obtain sufficient contrast from the fast-exchanging glutamate exchangeable protons than is needed for slow-exchanging aliphatic rNOE protons\cite{van2018magnetization}. Therefore, a relatively long scan time is needed for the acquisition of multi-metabolite CEST-weighted information.

% not sure about "The use of CEST MRF to provide quantitative parameter maps of..." sentance
Magnetic resonance fingerprinting (MRF) is an increasingly recognized technique for quantitative imaging\cite{ma2013magnetic}. The technique uses a pseudorandom, rapid pulse sequence to generate unique tissue-specific signal trajectories, followed by a pattern-matching algorithm, which assigns each voxel to a particular combination of T$_1$, T$_2$, and proton density values\cite{poorman2020magnetic}.
Following the successful application in water relaxometry\cite{chen2016mr, chen2019three}, MRF was subsequently expanded and modified for additional contrast mechanisms\cite{kratzer20213d, goren2025harnessing, fan2024simultaneous}. The use of CEST MRF to provide quantitative parameter maps of proton exchange rate (k$_{sw}$) and volume fraction (f$_{s}$)\cite{perlman2023mr} was first reported in 2018\cite{Cohen2018, zhou2018chemical, heo2019quantifying}. Although this approach was successful in quantifying a variety of biochemical targets, such as the exchangeable protons of Iohexol\cite{Perlman2022AutoCEST}, creatine\cite{liu2022encoding, Perlman2022AutoCEST}, and the aliphatic relayed nuclear rNOE\cite{power2024vivo}, most studies involved in vitro phantoms\cite{perlman2020cest}, wild type mice\cite{nagar2023dynamic}, or healthy human volunteers\cite{finkelstein2025multi, power2024vivo}. To date, all preclinical/clinical applications involving CEST MRF have focused on amide proton-based tumor characterization\cite{cohen2023cest, weigand2023accelerated, singh2025saturation} and cancer treatment monitoring\cite{perlman2022quantitative, vladimirov2023molecular}.

\begin{figure}[ht]
\includegraphics[width=1\textwidth]{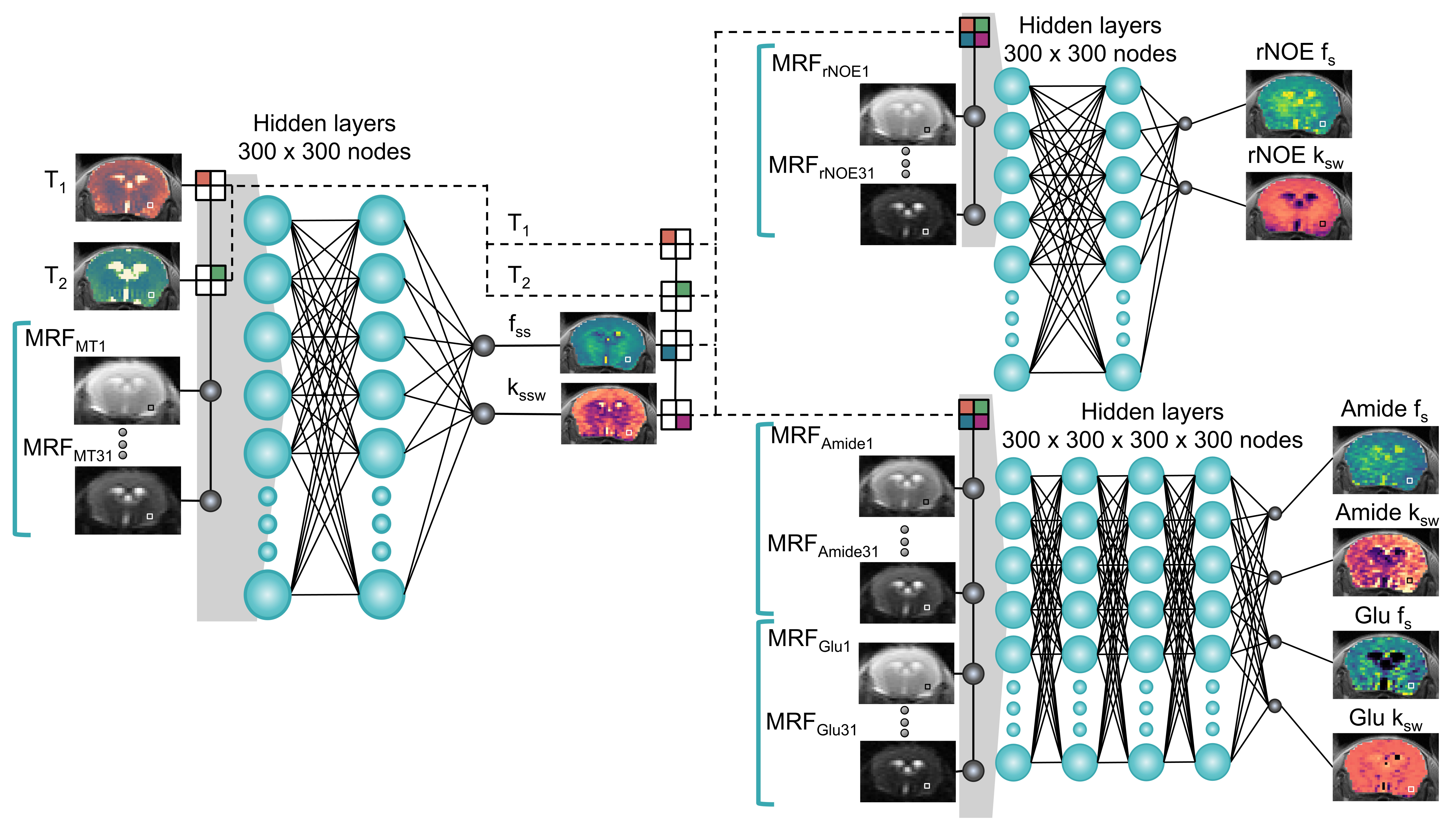}
\caption{\textbf{A deep learning pipeline for multi-metabolite chemical exchange saturation transfer magnetic resonance fingerprinting (CEST MRF).} Four rapid pulse sequences are implemented (Supplementary Fig. 1), where each yields 31 raw molecular-feature-embedded images for the semisolid magnetization transfer (MRF$_{MT}$), aliphatic relayed nuclear Overhauser effect (MRF$_{rNOE}$), amide (MRF$_{amide}$) and glutamate (MRF$_{Glu}$) proton pools (shown in gray-scale). These images, together with the water T$_1$ and T$_2$ maps (color coded brains, left), then serve as input for three fully connected neural networks, applied sequentially and in a pixelwise manner. Relevant information is shared across the pipeline (squared input neurons) to improve the quantification accuracy. The ultimate pipeline output are eight proton volume fraction (f$_{ss}$ / f$_{s}$) and exchange rate (k$_{ssw}$ / k$_{sw}$) maps for the semisolid MT or rNOE, amide, and glutamate proton pools, respectively.}
\label{fig:architecture}
\end{figure}

% changed utilized to validated
Here, we expand CEST MRF into a unified multi-metabolite imaging pipeline (Fig. \ref{fig:architecture}), designed to quantify and characterize a range of compounds that may play a role in PD. This non-invasive approach serially employs four different short pulse sequences (Supplementary Fig. 1) that encode the semisolid magnetization transfer (MT), amide, glutamate, and rNOE information into unique signal trajectories\cite{perlman2022quantitative, perlman2020cest, power2024vivo}. These are then subjected to quantitative parameter decoding by a series of three fully connected artificial neural networks (NN)\cite{perlman2022quantitative, vladimirov2025quantitative}, trained on synthetic data (Fig. \ref{fig:architecture}). We validated this imaging pipeline by characterizing the proton exchange parameters in a 1-methyl-4-phenyl-1,2,3,6-tetrahydropyridine (MPTP) mouse model of PD and comparing the results to those obtained from ground truth histology, traditional CEST-weighted imaging, and $^1$H MRS findings. 

% fig1 fig1_all.jpg

\section*{Results}

\subsection*{CEST MRF quantification in vitro}
\begin{figure}[ht]
\includegraphics[width=1\textwidth]{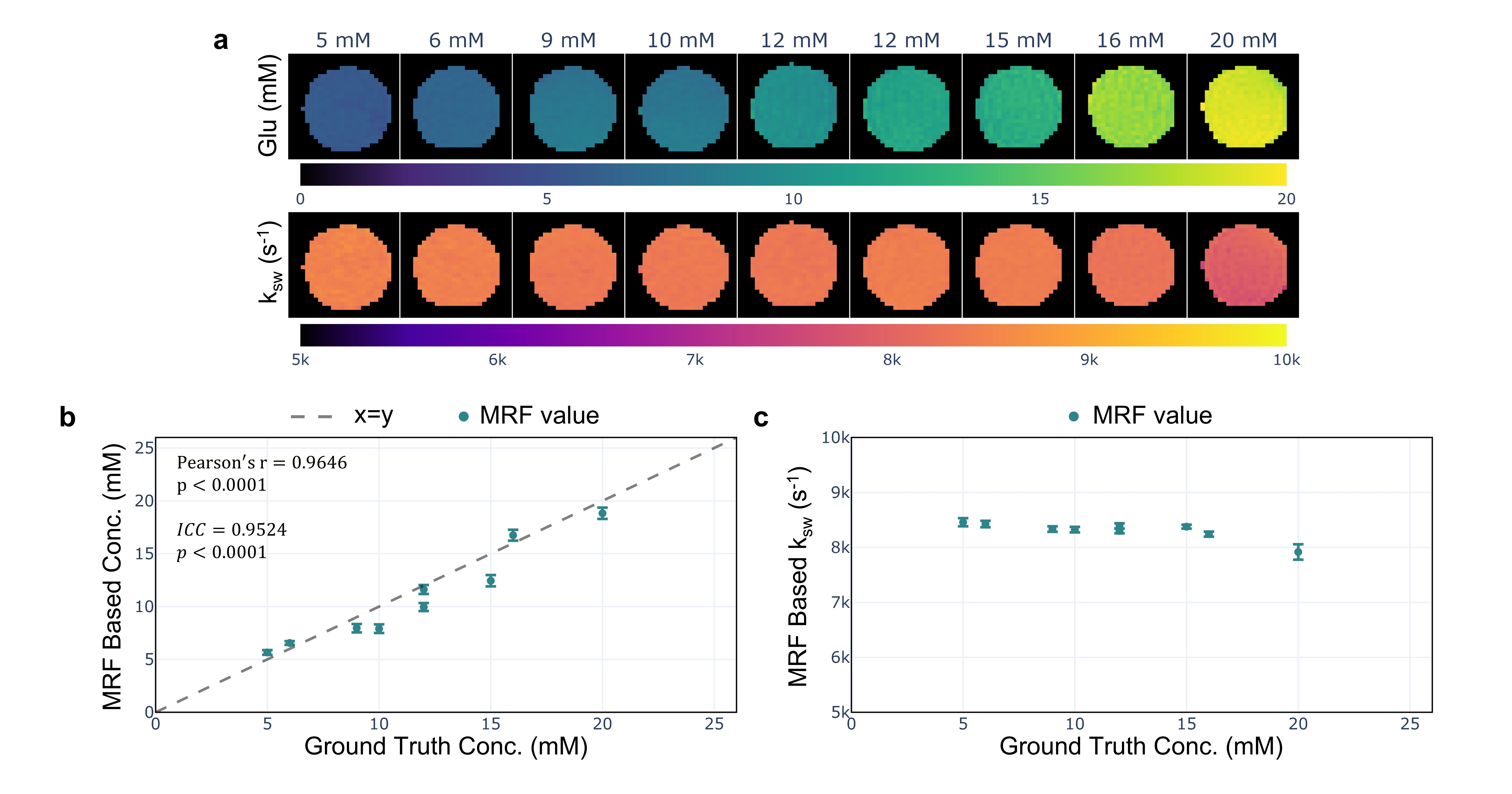}
\caption{\textbf{In vitro quantification of glutamate proton exchange parameters.} \textbf{a.} MRF-based glutamate concentration (top) and proton exchange rate (bottom) maps, acquired under physiological conditions (pH = 7.0, T = 37 °C) at 7T. \textbf{b.} MRF-determined glutamate concentrations in phantoms correlated significantly with the known concentrations (Pearson’s r = 0.9646, p < 0.0001, n = 9 vials). \textbf{c.} The MRF determined proton exchange rates for all phantoms were successfully decoupled from the concentration dynamics. Data are presented as mean $\pm$ standard deviation.}
\label{fig:phantoms}
\end{figure}

To the best of our knowledge, although CEST MRF has previously been used to quantify the semisolid MT, amide, and rNOE proton exchange parameters in vitro 
\cite{Singh2023, perlman2022quantitative, power2024vivo}, there has been no MRF-based quantification of glutamate under physiological conditions. In view of this gap, we prepared nine phantoms containing biologically relevant glutamic acid concentrations (5 to 20 mM, pH 7.0) as a preliminary step toward in vivo characterization. The phantoms were imaged at 37 °C using a preclinical 7T scanner (Bruker, Germany) and a spin echo echo-planar imaging (SE-EPI) MRF protocol (Supplementary Fig. 1)\cite{perlman2020cest}. Excellent agreement was obtained between the MRF-based concentrations and the ground truth (Fig. \ref{fig:phantoms}b), as reflected by a strong correlation (Pearson's r = 0.9646, p < 0.0001, n = 9 vials) and high reliability (intraclass correlation coefficient (ICC) = 0.9524, p < 0.0001, n = 9 vials). The MRF-based proton exchange rates were relatively stable at 8,307 $\pm$ 152 $s^{-1}$ and were successfully decoupled from the concentration dynamics (Fig. \ref{fig:phantoms}c). The measured exchange rate was in general agreement with previously reported (non-MRF) in vitro values\cite{Wermter2015,Khlebnikov2019}.

% fig2

\subsection*{In vivo CEST MRF in a PD mouse model}
\begin{figure}[ht]
\includegraphics[width=1\textwidth]{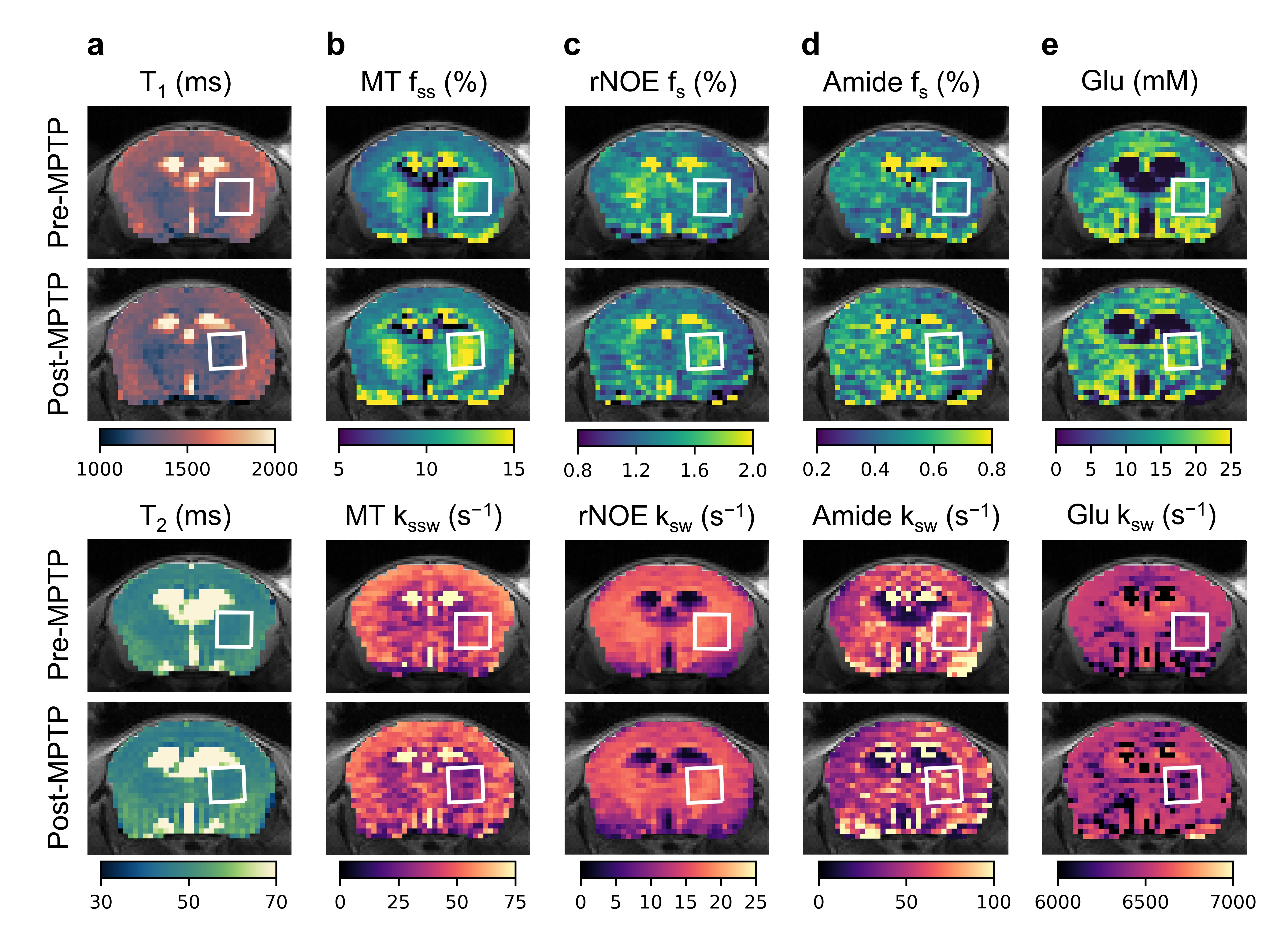}.
\caption{\textbf{Quantitative molecular images and relaxometry of a representative mouse before and after MPTP administration.} \textbf{a.} T$_1$ and T$_2$ relaxometry maps. \textbf{b.} Semisolid MT proton volume fraction (f$_{ss}$, top) and exchange rate (k$_{ssw}$, bottom) maps. \textbf{c.} rNOE proton volume fraction ($f_s$, top) and exchange rate ($k_{sw}$, bottom) maps. \textbf{d.} Amide proton volume fraction ($f_s$, top) and exchange rate ($k_{sw}$, bottom) maps. \textbf{e.} Glutamate concentration (top) and amine proton exchange rate ($k_{sw}$, bottom) maps. White squares represent the striatal region of interest (ROI). Note the T$_1$ relaxation shortening and elevated semisolid MT, amide, and rNOE proton volume fractions post MPTP administration.}
\label{fig:hero}
\end{figure}

\begin{figure}[ht]
\begin{center}
\includegraphics[width=0.76\textwidth]{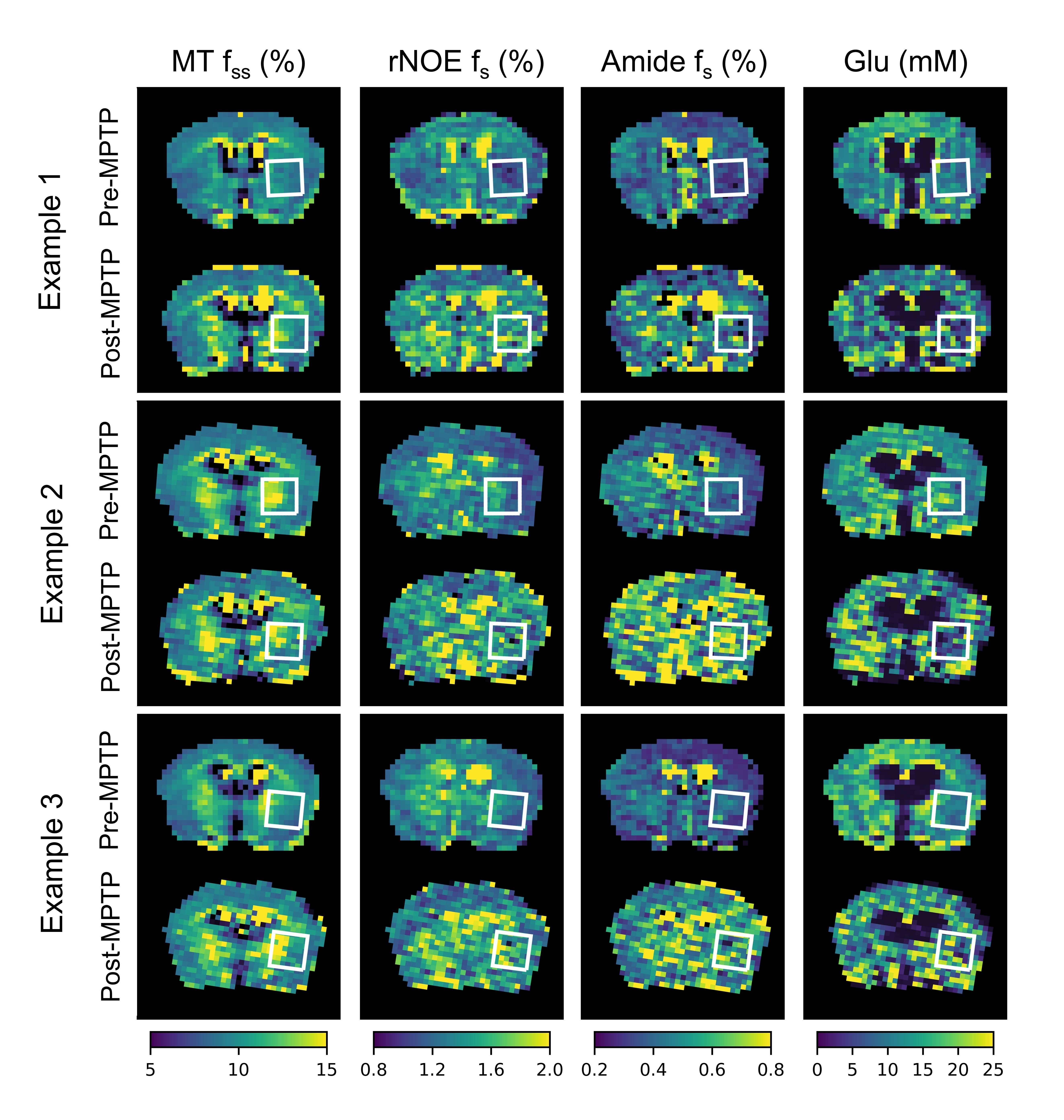}.
\end{center}
\caption{\textbf{Proton volume fraction maps from three additional representative mice, before and after MPTP administration}. Note the increased semisolid MT, amide, and rNOE proton volume fraction in the striatum (white squared region of interest).}
\label{fig:3reps}
\end{figure}

\begin{figure}[ht]
\includegraphics[width=1\textwidth]{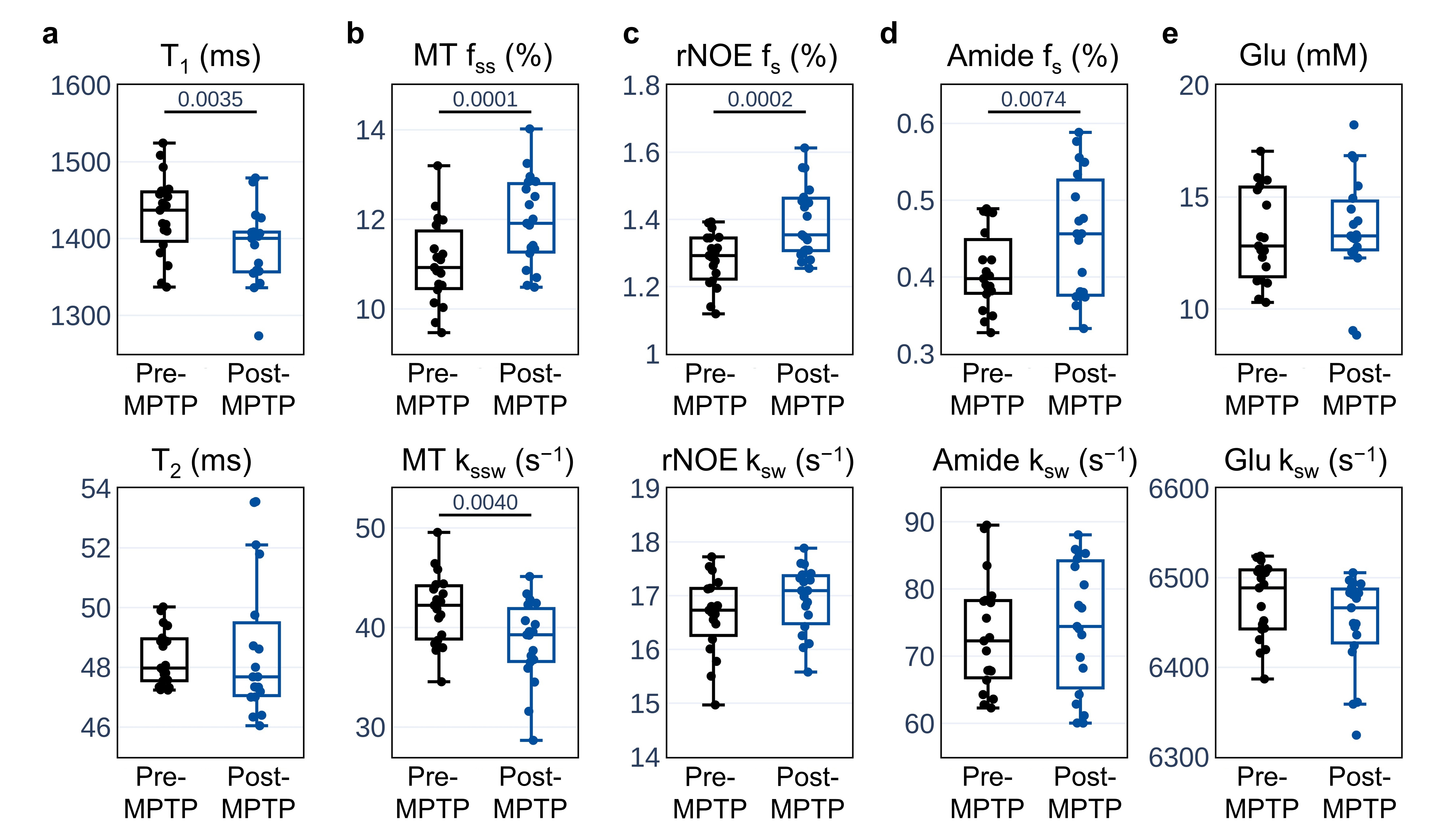}
\caption{\textbf{Statistical analysis of the CEST MRF-based parameters, before and after MPTP administration.} The analysis was performed on the striatum region of interest (ROI), as delineated in Fig.\ref{fig:hero} and \ref{fig:3reps}. \textbf{a.} Water T$_1$ and T$_2$ relaxation times. \textbf{b,c,d.} Proton volume fraction (top) and exchange rate (bottom) for the semisolid MT, rNOE and amide proton pools, respectively. \textbf{e.} Glutamate concentration (top) and amine proton exchange rate (bottom). The central horizontal lines in all box plots mark the median, the box limits represent the upper (third) and lower (first) quartiles, the whiskers represent 1.5 $\times$ the interquartile range above and below the upper and lower quartiles, respectively, and all data points are plotted. 
}
\label{fig:stats}
\end{figure}

The MPTP mouse model (n = 19) was employed to analyze the multi-metabolite properties of PD in vivo. Animals were imaged at baseline and 7 days post-MPTP administration (acute dosing paradigm, see Methods section). The proton exchange parameter maps obtained for a representative mouse are shown in Fig. \ref{fig:hero}. The results indicate a shortening of the T$_1$ relaxation time following MPTP administration, and an increase in the volume fractions of the semisolid MT, amide, and rNOE proton pools in the striatum, compared to baseline. Although there was also an increase in striatal glutamate concentration post-MPTP in some animals (Fig. \ref{fig:hero}e), this effect was not consistent (see additional examples of pre- and post-MPTP proton volume fraction maps in Fig. \ref{fig:3reps}). The statistical analysis of the entire mouse cohort presented in Fig. \ref{fig:stats}, further supports the visually detectable proton exchange effects in the striatal region of interest (ROI). Specifically, there was a significant shortening in the T$_1$ relaxation time (t-test, p = 0.0035, n = 19), a decrease in the semisolid MT proton exchange rate (t-test, p = 0.0040, n = 19), and a simultaneous increase in the proton volume fraction of the amide, rNOE, and semisolid MT proton pools (t-test, p = 0.0074, p = 0.0002, p = 0.0001, respectively, n = 19). While glutamate concentrations showed a mild elevation trend post-MPTP, this effect was not statistically significant (t-test, p = 0.5347, n = 19, Fig. \ref{fig:stats}e).

% fig4

\subsection*{Histological validation}
The MRF-based molecular findings were compared to histology and immunohistochemistry (IHC) results from six mice (three random MPTP-treated mice and three age- and strain-matched untreated controls). Brains were fixed in paraformaldehyde, paraffin-embedded, and coronally sectioned to include the striatum (using the Allen Brain Atlas as reference\cite{lein2007genome}). Sections were stained with Coomassie blue for total protein content. IHC markers included antibodies against glutamate and the metabotropic glutamate receptor 3 (mGluR3) to evaluate glutamate distribution, antibodies against the glial fibrillary acidic protein (GFAP) to assess astrocytic activation, and 4',6-diamidino-2-phenylindole (DAPI) for nuclear counterstaining. All imaging procedures used the same uniform conditions. Representative histological findings from an MPTP-treated mouse and a control mouse are presented in Fig. \ref{fig:hist_itay}. Additional examples are available in Supplementary Fig. 2. The clear increase in GFAP staining seen in MPTP-treated mice (Fig. \ref{fig:hist_itay}d, Supplementary Fig. 2a), reflects astrocytic activation, which is a hallmark of PD neuroinflammation\cite{kohutnicka1998microglial}. The Coomassie blue-stained image revealed an overall increase in protein concentration in the striatum compared to non-treated mice. Taken together, the GFAP and Coomassie blue findings are consistent with the CEST MRF findings of increased rNOE and amide proton volume fractions (Figs. \ref{fig:hero}, \ref{fig:3reps}, \ref{fig:stats}, Supplementary Fig. 3), and are indicative of elevated bulk mobile protein content\cite{goerke2018dual}. 

There was an increase in anti-glutamate staining in the striatum following MPTP administration (Fig. \ref{fig:hist_itay}a), which is consistent with the glutamatergic dysregulation commonly observed in PD\cite{mustapha2021mptp}. The specificity of this signal, was confirmed by co-localization of staining with the glutamate receptor mGluR3 signal (Fig. \ref{fig:hist_itay}b,c). These findings support the elevated glutamate concentrations detected by CEST MRF (Fig. \ref{fig:hero}e). Notably, there was a more modest increase glutamate staining in certain MPTP-treated mice compared to baseline (Supplementary Fig. 2c), which could explain the somewhat inconsistent CEST MRF detected increase in glutamate (Fig. \ref{fig:stats}e). A merged image summarizing the IHC results, presented in Fig.\ref{fig:hist_itay}e, illustrates the variety of molecular and cellular changes associated with MPTP-induced neurodegeneration.

% fig5
\begin{figure}[ht]
\includegraphics[width=1\textwidth]{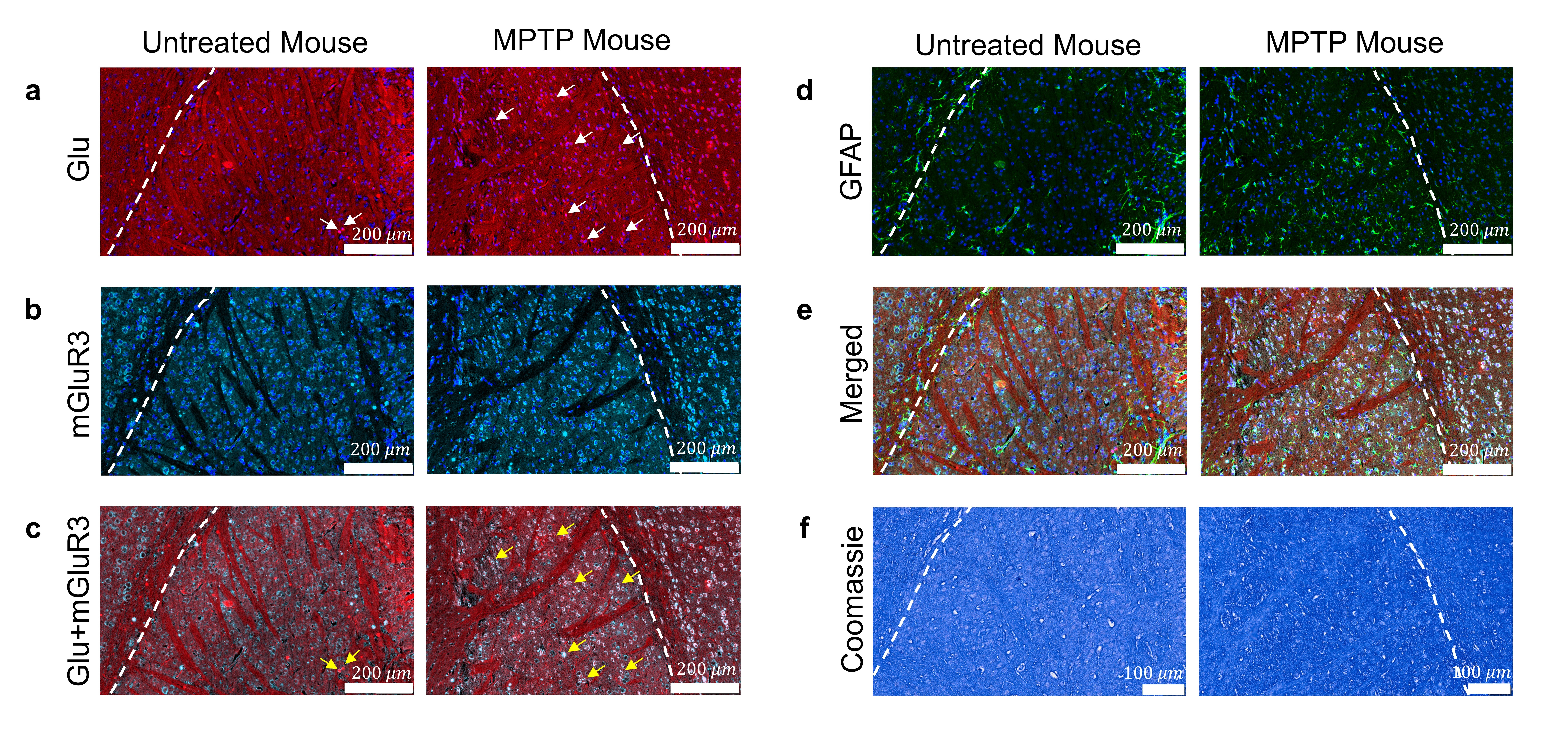}
\caption{\textbf{Histological analysis comparing a representative untreated mouse with a representative MPTP-treated mouse.} The white dashed lines in all panels mark the cortex-striatum boundary. \textbf{a.} Glutamate (Glu) and DAPI staining. The white arrows highlight Glu-positive signals in the striatum, presenting an increase after MPTP administration. \textbf{b.} Glu receptor (mGluR3) and DAPI staining. \textbf{c.} Glu and mGluR3 receptor staining. The yellow arrows indicate double-stained areas in the striatum, demonstrating enhanced expression in the MPTP-treated mouse. \textbf{d.} GFAP and DAPI staining, with an increased GFAP signal in the MPTP group, reflecting higher astrocytic activation (with DAPI providing background cell body visualization). \textbf{e.} A merged image combining the signals from panels \textbf{a}, \textbf{b}, and \textbf{d}. \textbf{f.} Coomassie blue staining. The increased staining intensity in the MPTP-treated brain suggests an increase in total protein content.}
\label{fig:hist_itay}
\end{figure}

\subsection*{A comparative analysis with CEST-weighted imaging and MRS}
The CEST MRF-based biomarkers were compared to those from classical CEST-weighted imaging, by acquiring two full Z-spectra with typical parameters for either slow- or fast-exchanging proton pools (see Methods section). The results indicated a significant increase in the striatal magnetization transfer ratio (MTR) signals, associated with the rNOE protons at -3.5 ppm and the amide protons at 3.5 ppm, after MPTP treatment (t-test, p = 0.0246 and p = 0.0463, n = 19, Supplementary Fig. 4a,b, respectively). This is in agreement with the increased proton volume fraction of these compounds quantified by CEST MRF (Fig. \ref{fig:stats}). However, these differences essentially disappear when the MTR values are fused into the traditional MTR$_{asym}$(3.5 ppm) metric (t-test, p = 0.9698, n = 19, Supplementary Fig. 4c), because of mutual compensation between the two signals.

Similarly, GluCEST-weighted imaging did not detect any statistically significant difference between the striatal MTR$_{asym}$(3 ppm) values of the baseline and MPTP groups (t-test, p = 0.6078, n = 11, Supplementary Fig. 4e). To further investigate the detectability of glutamate via non-CEST MR, $^1$H MRS was used in all mice, both before and after MPTP treatment. The results identified a general glutamate concentration increase following treatment (Supplementary Fig. 4d), but this effect did not reach statistical significance (t-test, p = 0.0547, n = 19).

Lastly, we used an amine and amide concentration independent detection (AACID) method\cite{McVicar2014} to assess potential intracellular pH alterations in the striatum. The results did not identify any statistically significant effects of treatment (t-test, p = 0.4418, n = 19, Supplementary Fig. 4f). This is in agreement with the lack of statistically significant differences in the proton exchange rates of amide and glutamate, before and after MPTP treatment, as quantified by CEST MRF (Fig. \ref{fig:stats}).

\section*{Discussion}

% general PD info
PD is a highly disruptive and common neurodegenerative disease that afflicts approximately 0.15\% of the global population\cite{Zhu2024}. Diagnosis typically relies on the presence of motor symptoms, which only become apparent after substantial disease progression\cite{Kordower2013}. Consequently, there is an urgent need for new methods that can facilitate earlier and more accurate diagnosis and distinguish between disease subtypes. Molecular MRI offers a non-invasive means for probing PD-related biochemical changes without the use of ionizing radiation. In this context, MRS has shown particular promise\cite{ciurleo2014magnetic} but is hindered by low sensitivity, resulting in long acquisition times and limited spatial resolution. CEST imaging offers improved spatial resolution\cite{liu2013nuts}, but provides semi-quantitative information, and traditionally required long acquisition times for quantification studies\cite{ji2023vivo}. The development of CEST MRF has introduced new opportunities for rapid quantification of molecular properties in vivo\cite{cohen2023cest, weigand2023accelerated, singh2025saturation,perlman2022quantitative, vladimirov2023molecular}.

Here, we further leveraged CEST MRF and AI to develop a unified framework designed to characterize a number of key metabolites and compounds related to the molecular and cellular pathways associated with PD. We implemented a 'divide and conquer' approach, in which a series of rapid pulse sequences are applied separately\cite{perlman2022quantitative}, and where each encodes a specific proton pool, while minimizing the contributions made by other proton pools (Supplementary Fig. 1). Since the semisolid MT signal affects the entire Z-spectrum, the parameter quantification pipeline starts by extracting the proton exchange parameters of this pool, using the information obtained from a pulse sequence focused on a spectral range where no other metabolites are "visible", and taking the water T$_1$ and T$_2$ relaxation times into account (Fig. \ref{fig:architecture}). The pipeline then diverges into two separate neural networks, which quantify the rNOE or amide/glutamate proton exchange parameters. The rationale behind the simultaneous quantification of amide and glutamate is the need to account for any mutual signal overlap resulting from the spectral proximity of their chemical shifts. In all cases, the information gained during the previous steps is shared with the subsequent neural networks (Fig. \ref{fig:architecture}). 

The preliminary results of our in vitro study demonstrated that glutamate concentrations can be quantified under physiological conditions using CEST MRF (Pearson's r = 0.9646, p < 0.0001, n = 9, Fig. \ref{fig:phantoms}). While the associated proton exchange rates were successfully decoupled from the concentration dynamics, they were slightly higher than previous in vitro reports (7,248 $\pm$ 189 $s^{-1}$\cite{Wermter2015}, 7,480 $\pm$ 90 $s^{-1}$\cite{Khlebnikov2019}). This could be attributed to fan-generated, uneven over-heating of the phantom vials. 

In vivo, CEST MRF-based glutamate concentration maps revealed a visible striatal increase post-MPTP treatment (Fig. \ref{fig:hero}e), yet this effect was inconsistent across the entire mouse cohort (Fig. \ref{fig:stats}). Similarly, the IHC results (Fig. \ref{fig:hist_itay}, Supplementary Fig. 2,3) identified an elevation of glutamate in certain animals, but not others. Notably, while early CEST-weighted studies reported an increase in the glutamate signal in an MPTP mouse model\cite{bagga2016mapping, Bagga2018}, there were no statistically significant differences detected here after applying the same GluCEST-weighted approach (t-test, p = 0.6078, n = 11, Supplementary Fig. 4e). Moreover, while $^1$H MRS quantification detected a general increase in striatal glutamate concentration after MPTP exposure (Supplementary Fig. 4d), the differences did not reach statistical significance (t-test, p = 0.0547, n=19). Importantly, successful implementation of the MPTP model was confirmed by consistent GFAP elevation, along with observed physiology and motor symptoms (including weight loss, muscle rigidity, and, in some cases, seizures) in the treated mice. We suggest that the discrepancy between our glutamate-related results and those of previous GluCEST-weighted studies\cite{bagga2016mapping, Bagga2018} may be attributed to differences in the MPTP administration regimen. Specifically, our current study employed an acute administration route\cite{jackson2007protocol}, as opposed to a chronic injection regimen used previously\cite{bagga2016mapping, Bagga2018}. Acute MPTP models induce rapid neurotoxicity and dopaminergic neuron loss, but may not allow sufficient time for metabolic adaptations such as glutamate accumulation to develop uniformly\cite{Qi2023, Klemann2015}. 

In contrast to the inconclusive glutamate findings, the CEST MRF proton volume fractions maps of the semisolid MT, amide, and rNOE revealed a clear (Figs. \ref{fig:hero}, \ref{fig:3reps}) and statistically significant (Fig. \ref{fig:stats}) increase in the striatum. A potential explanation for these effects is the pronounced inflammatory response characteristic of PD, particularly in the acute MPTP model\cite{KurkowskaJastrzbska2009}. This response includes astrogliosis, microglial activation, infiltration of immune cells, and elevated pro-inflammatory cytokines in the striatum\cite {Luchtman2009, KurkowskaJastrzbska2009}. Astrogliosis and microglial activation probably contribute to increased glial membrane density and cytoskeletal protein content. The increased amide proton volume fraction can be attributed to the accumulation of cytosolic proteins and peptides, resulting from neuronal degeneration, increased protein synthesis by reactive glia, and infiltrating immune cells. Similarly, the observed increase in the aliphatic rNOE proton volume fraction may reflect changes in membrane lipid composition associated with glial activation and immune cell infiltration\cite{Zu2020, Brekk2020}.

Histological analysis provided further support for the proton volume fraction trends revealed by CEST MRF. Specifically, there was a considerable increase in the signal intensity of GFAP IHC staining in MPTP-treated mice, which is consistent with the presence of astrogliosis (Fig. \ref{fig:hist_itay}d and Supplementary Fig. 2a). Coomassie blue staining confirmed a general increase in total protein content (Fig. \ref{fig:hist_itay}f and Supplementary Fig. 2b), in agreement with the increase in bulk mobile protein content demonstrated in the CEST MRF amide and rNOE proton volume fraction maps.

Finally, we compared the CEST MRF findings with CEST-weighted analysis of the same mice and ROIs. The results revealed statistically significant increases in the raw MTR values, associated with the rNOE and amide proton pools (at -3.5 ppm and 3.5 ppm, respectively) post MPTP treatment (t-test, p = 0.0246, p = 0.0463, respectively, n = 19, Supplementary Fig. 4a,b). However, this significance disappeared when these two metrics were combined into the conventional MTR$_{asym}$ formula, (pre- and post-MPTP treatment t-test, p = 0.9698, n = 19, Supplementary Fig. 4c). This effect underscores the benefits of separating and distilling the different biophysical properties of each proton pool, as performed in CEST MRF. 

In addition to the proton volume fraction dynamics post MPTP treatment, the results of our CEST MRF study also detected a shortening of the water T$_1$ (t-test, p = 0.0035, n = 19, Fig. \ref{fig:stats}) in agreement with previous relaxometry studies in PD\cite{keil2020pilot, Nrnberger2017}. Furthermore, the semi-solid proton exchange rate was decreased following MPTP (t-test, p = 0.0040, n = 19, Fig. \ref{fig:stats}). This may be attributed to a change in lipid composition of the cell membranes compared to healthy brain tissue, which alters the base-catalysed exchange rate constant of the semi-solid protons\cite{perlman2022quantitative}.

The acquisition time required for all CEST MRF acquisition protocols (Supplementary Fig. 1) was 8.25 minutes. While the incorporation of T$_1$ and T$_2$ maps increases the overall scan time, they can be acquired in less than 0.5 minutes using traditional water-pool MRF\cite{ma2013magnetic}. Therefore, when combined with the very short NN inference (less than 1 s), CEST MRF provides a rapid means for the quantitative assessment of molecular PD information. Notably, all the MRF pulse sequences used here were based on previous CEST MRF studies and their insights\cite{perlman2022quantitative, Cohen2018, power2024vivo}. Future work will leverage classical optimization strategies\cite{vladimirov2025optimization} and AI-based protocol discovery frameworks\cite{Perlman2022AutoCEST, Kang2021loas} to reduce scan time even more, while maintaining, or improving, the accuracy of parameter quantification.

\subsection*{Conclusions}
This study describes a unified molecular imaging framework for rapid and quantitative characterization of multiple metabolites and compounds in vivo. Our findings highlight the potential of CEST MRF as a valuable tool that can detect molecular alterations in PD. The modular design of the acquisition and reconstruction pipeline enables a flexible implementation tailored to scan time constraints (e.g., by choosing a subset of molecular targets to focus on), with the potential for adaptations for additional neurological disorders. 

\section*{Methods}

\subsection*{Phantom preparation}
Nine 2 ml vials were filled with glutamic acid (glutamate, amine protons with a chemical shift = 3 ppm with respect to water, Sigma-Aldrich) at concentrations of 5-20 mM dissolved in 10 mM phosphate buffered saline (PBS) and titrated to pH 7.0.

\subsection*{Animal preparation}
All experimental protocols adhered to the ethical principles of the Israel National Research Council (NRC) and received approval from the Tel Aviv University Institutional Animal Care and Use Committee (IACUC) (TAU-MD-IL-2309-160-5). Male C57BL/6 mice (8 weeks old, n = 22) were acquired from Harlan Laboratories (Israel). The experimental group (n = 19) received four intraperitoneal (IP) injections of 1-methyl-4-phenyl-1, 2, 3, 6-tetrahydropyridine (MPTP-HCl, dissolved in saline, CAS 23007-85-4, BioTAG), at a dose of 14-20 mg/kg, administered every 2 hours over an 8 hour period. This treatment is expected to yield 40\% or greater striatal dopamine depletion at 7 days post injection\cite{mustapha2021mptp}. Three mice served as control for the histological analysis and were not subjected to the MPTP injection protocol.

\subsection*{MRI acquisition}
All imaging studies were performed using a preclinical 7T scanner (Bruker, Germany). Phantoms were imaged at 37 °C, maintained using a hot air blower (SA Instruments, NY, USA). Animals were scanned at baseline (pre-MPTP treatment), and at 7 days post treatment. The animals were anesthetized with 0.5-2\% Isoflurane throughout the scan, with the respiration rate supervised via a physiological monitoring system (SA Instruments). The imaging cradle included a hot water circulation system that maintained the body temperature of the animal at 37 °C. 

Four different CEST MRF protocols were sequentially applied to encode the semisolid MT, amide, rNOE, and glutamate information into unique signal trajectories. Each protocol acquired 31 raw images, the first being an M$_0$ image with repetition time and echo time (TR/TE) = 15,000/20 ms. The exact saturation pulse properties used for all protocols are described in supplementary Fig. 1\cite{perlman2022quantitative, Cohen2018, power2024vivo, vladimirov2025quantitative}, and required a scan time of 120 s to 135 s per protocol (8.25 min for the entire MRF acquisition). T$_1$ maps were acquired in vivo using the rapid acquisition with relaxation enhancement (RARE) protocol, with TR = 200, 400, 800, 1,500, 3,000, and 5,500 ms, TE = 7 ms, RARE factor = 2, acquisition time = 364.8 s. T$_2$ maps were acquired using the multi-echo spin-echo (MSME) protocol, TR = 2,000 ms, with 25 TE values between 20 and 500 ms and acquisition time = 128 s. B$_0$ maps were acquired using the water saturation shift referencing (WASSR) protocol\cite{Kim2009}, with a saturation pulse power of 0.3 $\mu$T, TR/TE = 8,000/20 ms, FA = 90°, saturation duration = 3,000 ms, saturation pulse frequency offset varying between -1 ppm and 1 ppm with 0.1 ppm increments and acquisition time = 176 s.

% some rephrasing needed
Three full Z-spectra CEST scans were performed, in order to obtain slow-exchange amide/rNOE CEST-weighted information\cite{perlman2022quantitative}, fast-exchanging glutamate amine-weighted information\cite{bagga2016mapping, Bagga2018}, and pH-weighted AACID information\cite{McVicar2014}. The saturation pulse powers were 0.7/5.9/1.5 $\mu$T, the saturation pulse durations were 3/0.5/4 s, and the acquisition times were 928/928/1,152 s for the slow-exchange/fast-exchange/AACID oriented schedules, respectively. The frequency offset ranged between 7 to -7 ppm in all cases, with the frequency increments being 0.25/0.25/0.2 ppm, for the slow-exchange/fast-exchange/AACID oriented schedules, respectively. In all cases a SE-EPI readout was employed, with a FA = 90°, TE = 20 ms, TR = 8,000 ms, and two averages.

The field of view (FOV) was 32 $\times$ 32 $\times$ 5 mm$^3$ in vitro and 19 $\times$ 19 $\times$ 1.5 mm$^3$ in vivo. The in-plane resolution was set to 500 $\mu$m and 297 $\mu$m for the in vitro and in vivo cases, respectively. A T$_2$-weighted scan (TR/TE = 2,000/60 ms) with an in-plane resolution of 148 $\mu$m and 1.5 mm slice thickness was taken as reference. 

$^1$H-MRS spectrum was measured from a 2 $\times$ 2 $\times$ 1.5 mm$^3$ voxel at the striatum in all mice using the Point RESolved Spectroscopy (PRESS) sequence with 128 water-suppressed signal averages and 8 signal averages without water suppression (at 4.7 ppm, for water reference signal normalization). Water suppression, calibration, and shimming were performed by an automatic pre-scanning procedure of each voxel. The full measurement lasted 1,088 s. The sequence parameters were TR = 8,000 ms, TE = 16 ms, spectral width = 4,000 Hz, fidres = 0.4 Hz. The total acquisition time for a complete single mouse scan was 1.5 h, including all comparative scans and MRS.

\subsection*{MRF dictionary generation} 
Simulated signal trajectories were generated using our open-access numerical solver for the Bloch–McConnell equations\cite{vladimirov2025quantitative}, implemented in C++ according to the Pulseq definition standard\cite{herz2021pulseq}, with a Python front-end that offers parallelization capabilities. This process yielded a total of 216,759,522 simulated signals for various tissue parameter combinations. Generating all dictionaries used in this work required a total of 6.5 h, using a computing server employing 50-100 CPU workers. The detailed parameters of the simulated dictionaries are available in Supplementary Tables 1-4.

\subsection*{Deep learning quantification} 
Proton exchange parameter quantification in vivo was performed in a pixelwise manner\cite{perlman2022quantitative, cohen2018mr}, using a series of three fully connected NNs (Fig. \ref{fig:architecture}). First, MT-oriented signal trajectories were input into the first network, together with the pixelwise water T$_1$ and T$_2$ values, which allowed us to quantify the semisolid MT proton volume fraction and exchange rate. These two parameters, together with the water T$_1$ and T$_2$ relaxation values (squared input neurons in Fig. \ref{fig:architecture}), were input into two subsequent NNs. The aim of the first network was to quantify the rNOE proton exchange parameters\cite{power2024vivo}. The aim of the second network was to concomitantly quantify the amide and glutamate proton exchange parameters (while taking any mutual effects into consideration). Each network was fed with the appropriate raw CEST MRF encoding data following L$_2$ normalization, as described by the gray-scale images shown in Fig. \ref{fig:architecture}.
For the glutamate phantom study, image reconstruction utilized the same NN backbone as used in vivo. The input was the per-pixel Glu-specific MRF trajectories, and the output comprised the glutamate concentration and the glutamate amine proton exchange rate.
All NNs included two hidden layers, except for the combined amide/glutamate NN in vivo, where there were four hidden layers because of the increased complexity. Each hidden layer consisted of 300 neurons. ReLU and Sigmoid activations were used for the hidden and output layers, respectively. Optimization was performed using the adaptive moment estimation (ADAM) algorithm\cite{Kingma2014} with a learning rate of 0.0002, a minibatch size of 1,024, and mean squared error as the loss function. White Gaussian noise was injected into the trajectories to promote robustness\cite{Zur2009}. The reconstruction pipeline was realized using Pytorch on an Intel(R) Xeon(R) Gold 655S CPU and an NVIDIA L40S GPU.

\subsection*{Histological analysis}
Three random MPTP-treated mice and three untreated control mice were prepared for histological analysis. Brains were harvested and immediately fixed in 4\% paraformaldehyde at 4°C for 24–48 hours, and then embedded in paraffin. 
Consecutive 5$\mu$m (coronal) sections were prepared to include the striatum area of interest. The sections were stained either by fluorescence immunohistochemistry (IHC) for specific markers or with Coomassie blue for protein content.

% comb-2 (what about dapi)
Antibodies against glutamate (\#AB5018, Sigma Aldrich), GFAP (\#Ab4674, Abcam), and mGluR3 glutamate receptor (\#AGC-010-GP, Alomone labs) were used for IHC tissue staining. Slides were deparaffinized and subjected to heat mediated antigen retrieval. After cooling, the slides were treated with blocking solution and incubated with the glutamate, GFAP and mGluR3 glutamate receptor antibody combination for 1 h at room temperature. After washing, the slides were incubated for 1 h with the species-specific secondary antibodies: donkey anti-rabbit Cy3 for glutamate (\#711-165-152, Jackson ImmunoResearch), donkey anti-chicken 488 for GFAP (\#703-545-155, Jackson ImmunoResearch), and donkey anti-guinea pig 647 for mGluR3 (\#706-605-148, Jackson ImmunoResearch). The slides were rinsed again and then counterstained with DAPI, washed, mounted, and coverslipped. After drying, the slides were scanned at 20$\times$ magnification using a fluorescence scanner (Olympus). All images for each antibody were acquired using the same exposure conditions.

Coomassie staining for protein concentration detection was performed as previously described\cite{Ray2019}. Slides were scanned at 20$\times$ magnification using a fluorescence scanner (Olympus).

\subsection*{MRS analysis}
TopSpin software (version 4.3, Bruker BioSpin GmbH, Germany) was used for spectral processing, where the free induction decays (FIDs) were applied a line-broadening factor of 5-10 Hz. All spectra were Fourier transformed, phased, baseline corrected, and calibrated to the N-Acetylaspartate (NAA) reference signal at $\delta$ = 2.02 ppm. Glutamate concentration was quantified using the unsuppressed water peak as an internal reference. The water signal, acquired from the same volume of interest (VOI), served as a concentration standard, assuming 80\% brain water content (approximately 44 mol/L)\cite{Tk2004}. Glutamate concentrations were quantified using the Topspin 4.3.0 software suite. The peak of the NAA served as internal reference for the glutamate concentration quantification, since under normal condition, the NAA concentration in the striatum is approximately 10 mM, based on MRS measurements\cite{Sager1999, Blml1999, Pan2004}, making it the most concentrated amino acid in the brain. Notably, no changes in NAA levels were reported in the MPTP mouse model\cite{Chassain2010, Petiet2021}.

\subsection*{CEST-weighted analysis}
All conventional CEST images were corrected for B$_0$ inhomogeneity using the WASSR method\cite{Kim2009, Liu2010}, and cubic spline smoothed\cite{Chen2013}. The magnetization transfer ratio (MTR) metric was calculated according to the formula: $MTR(\Delta\omega) = 1-S_{sat}(\Delta\omega) / S_0$\cite{Zhou2019}, where $S_{sat}(\Delta\omega)$ is the water signal during saturation at frequency offset $\Delta\omega$, and $S_0$ is the unsaturated signal. The magnetization transfer ratio asymmetry (MTR$_{asym}$) metric was calculated according to: $MTR_{asym}(\omega) = ({S_{sat}(-\omega)-S_{sat}(+\omega)})/ S_0$\cite{Zhou2019}. The Z-spectrum acquired with a saturation pulse power of 0.7 $\mu$T (see the MRI acquisition section) was used to calculate the rNOE-related MTR values at -3.5 ppm (Supplementary Fig. 4a), APT representing MTR values at 3.5 ppm (Supplementary Fig. 4b), and MTR$_{asym}$ values at $\pm$3.5 (Supplementary Fig. 4c).

The glutamate weighted signal (Supplementary Fig. 4e) was calculated based on the Z-spectrum with B$_1$ = 5.9 $\mu$T, as defined in previous GluCEST studies\cite{Bagga2018}: $GluCEST(\%) = (M_{sat} (-3 ppm)-M_{sat} (3 ppm))/(M_{sat} (-3 ppm)) \times 100$.

AACID values were calculated using the Z-spectra acquired with a saturation pulse power of 1.5 $\mu$T and the following formula\cite{McVicar2014}: $AACID = M_{sat}(3.5 ppm)\times(M_{sat}(6.0 ppm)-M_{sat}(2.75 ppm))/(M_{sat}(2.75 ppm)\times(M_{sat}(6.0 ppm)-M_{sat}(3.5 ppm)))$, where M$_{sat}$ = S$_{sat}$ / S$_0$.

\subsection*{Statistical analysis}
Pearson's r and ICC values (Fig. \ref{fig:phantoms}) were calculated using the SciPy\cite{virtanen2020scipy}, Scikit-Learn\cite{pedregosa2011scikit} and PyIRR\cite{gamer2010irr} libraries for Python. The circles in the error bars (Fig. \ref{fig:phantoms}), represent the mean and the bars represent the standard deviation (STD). In vivo striatal ROIs were delineated in the scanner during the data acquisition by a different researcher (M.R.) than the one who conducted the post-experiment MRI analysis (H.S.). The same ROIs were used for all CEST MRF, CEST-weighted, and MRS data analysis. The central horizontal lines in all the box plots, mark the median values, the box limits represent the upper (third) and lower (first) quartiles, the whiskers represent 1.5 $\times$ the interquartile range above and below the upper and lower quartiles, respectively, and all data points are plotted (Fig. \ref{fig:stats}, Supplementary Fig. 4). 
Paired two-tailed t-tests were calculated using the open-source SciPy library for Python\cite{virtanen2020scipy}. Statistical significance was set at p < 0.05.

\subsection*{Data availability}
All phantom and mouse data will become available upon acceptance at https://github.com/momentum-laboratory/multi-metabolite-pd and Zenodo. 

\subsection*{Code availability}
The code used in this work will become available upon acceptance at https://github.com/momentum-laboratory/multi-metabolite-pd and Zenodo. All CEST MRF acquisition protocols can be reproduced using the open MRI pulse sequences provided in https://osf.io/52bsg\cite{vladimirov2025quantitative}. 

%\bibliography{aux, mendeley}

\section*{Acknowledgements}
This work was supported by the Ministry of Innovation, Science and Technology, Israel. The authors thank Kai Geronik Sofer and Neta Zack for their
preliminary work and insights. This project was funded by the European Union (ERC, BabyMagnet, project no. 101115639). Views and opinions expressed are, however, those of the authors only and do not necessarily reflect those of the European Union or the European Research Council. Neither the European Union nor the granting authority can be held responsible for them. 

\section*{Author contributions statement}
H.S. and O.P. conceived the computational framework. M.R., H.S., and O.P. designed the phantom and animal studies. H.S. and M.R. acquired the imaging data. H.S. performed the AI design, optimization, and data analysis. M.R. analyzed the MRS data. H.S. wrote the manuscript. O.P. supervised the project. All authors reviewed and revised the manuscript.

\section*{Competing Interests}
The authors declare no competing interests.

\section*{Additional Information}
A supplementary information file is available.
\pagebreak

\end{document}